# *Quantum evolutionary algorithm for TSP combinatorial optimisation problem*

**YIJIANG MA [1], TAN CHYE CHEAH [2]**

1   YIJIANG MA; Department of Computer Science, University of Nottingham, Nottingham, United Kingdom
*   psxym11@nottingham.ac.uk
2   TAN CHYE CHEAH; School of Computer Science, University of Nottingham, Semenyih, Malaysia
    kcztcc@nottingham.edu.my

**Abstract**
This paper implements a new way of solving a problem called the traveling salesman problem (TSP) using quantum genetic algorithm (QGA). We compared how well this new approach works to the traditional method known as a classical genetic algorithm (CGA). The TSP is a well-established challenge in combinatorial optimization where the objective is to find the most efficient path to visit a series of cities, minimizing the total distance, and returning to the starting point. We chose the TSP to test the performance of both algorithms because of its computational complexity and importance in practical applications. We choose the dataset from the international standard library TSPLIB for our experiments. By designing and implementing both algorithms and conducting experiments on various sizes and types of TSP instances, we provide an in-depth analysis of the accuracy of the optimal solution, the number of iterations, the execution time, and the stability of the algorithms for both. The empirical findings indicate that the CGA outperforms the QGA in terms of finding superior solutions more quickly in most of the test instances, especially when the problem size is large. This suggests that although the principle of quantum computing provides a new way to solve complex combinatorial optimisation problems, the implementation of quantum phenomena and the setting of parameters such as the optimal angle for a quantum revolving gate is challenging and need further optimisation to achieve the desired results. Additionally, it is important to note that the QGA has not been tested on real quantum hardware, so its true performance remains unverified. These limitations provide rich opportunities for further research in the future.
**Keywords:** evolutionary algorithms; quantum computing; quantum algorithms; genetic algorithm; quantum genetic algorithm; traveling salesman problem

## 1 Introduction

Many researchers are increasingly interested in heuristic optimization algorithms, including genetic algorithms, Simulated Annealing, Ant Colony Optimization, Particle Swarm Optimization, and more. These methods are becoming a powerful tool for tackling NP problems. Quantum computing is a rising approach to computation, and its parallelism gives it a better ability to explore solutions when dealing with complex problems, so it is widely used in solving complex problems that require a large amount of computational space.

The research object of this paper is QGA. Firstly, we implemented the basic structure of QGA, then we used it to solve specific problems in experiments and compared its performance with CGA.

**1.1 Background**
This section shows the background related to quantum computing, evolutionary algorithm and quantum evolutionary algorithm.

**1.1.1 Quantum computing**
The quantum mechanical properties of quantum computing make it easy to solve some computational tasks that are difficult for classical computers. In quantum computing, information is stored in quantum bits, denoted as $|0\rangle$ and $|1\rangle$ states. Unlike classical computers, quantum bits appear in a superposition of 0 and 1 states or in a coherent state. The nature of quantum entanglement is such that the state of one quantum bit changes as the state of the other changes [1]. Quantum superposition states can be represented by the formula $|\psi\rangle = \alpha|0\rangle + \beta|1\rangle$ ($|\rangle$ is the Dirac notation) where $\alpha$ and $\beta$ denote the qubit amplitudes of the states 0 and 1, respectively. $|\alpha|^2$ and $|\beta|^2$ represent the likelihood of a quantum bit being in the 0 and 1 states respectively, so $|\alpha|^2 + |\beta|^2 = 1$ [2].

A quantum gate is a fundamental operation in quantum computing. It is a special matrix that can act on the state vectors of quantum bits, which results in new state vectors that can change the state of one or more quantum bits.

Hadamard Gate is a very important quantum gate in quantum computing, which acts on a single quantum bit. When a quantum bit has a base state of $|0\rangle$ or $|1\rangle$, the H-gate can transform it into a superposition state, when the quantum bits take on the value 0 or 1 with equal probability [3]. This property of superposition state enables quantum bits to perform operations with the superposition property in the concept of quantum mechanics. This represents a key distinction between quantum computing and classical computing.

Pauli gates are important single-qubit gates in quantum computing, including the Pauli-X, Pauli-Y and Pauli-Z gates. The Pauli X Gate is also known as a bit-flip gate or NOT gate. It acts similarly to a logical NOT gate in a classical computer by flipping the states of $|0\rangle$ and $|1\rangle$. The Pauli-Y Gate alters the quantum bit's state and simultaneously introduces a complex phase represented by the imaginary unit 'i'. In quantum computation, information is not only stored in the ground state, but is also contained in the phase. Quantum computing can use both the ground state and the phase to process information, which is more efficient than classical computing. Pauli Z Gate does not change the $|0\rangle$ state but adds a negative phase to the $|1\rangle$ state, and this clever use of the phase can achieve some functions that classical computing cannot [4].

Other common quantum gates include CNOT gates, Toffoli gates and Fredkin gates. CNOT gates and Toffoli gates are Controlled gates. They are a special class of operations on multiple quantum bits in quantum computing. This operation determines whether to execute a specific action on the second quantum bit, contingent upon the state of the first quantum bit. For a CNOT gate, a Pauli X gate is applied to the target quantum bit when the control quantum bit is in the $|1\rangle$ state. Toffoli gate is a three-qubit gate, where the program performs a NOT operation on the third target quantum bit when the first two control quantum bits are in the $|1\rangle$ state. Fredkin gate, also known as a controlled SWAP gate, is a three-bit quantum gate. If the first control bit is in the $|1\rangle$ state, then it exchanges the states of the next two quantum bits; if the first control bit is in the $|0\rangle$ state, then the states of the next two quantum bits remain unchanged [2].

Quantum computing utilises the superposition nature of quantum mechanics can allow computers to increase their computational power dramatically. The application of quantum computing to Grover's search algorithm allows for a quadratic polynomial time speedup [3]. Quantum computing also has advantages in machine learning. When faced with complex data in high-dimensional spaces, quantum computing can efficiently process these data and discover relationships between features, which is not possible with classical computers. By comparing the results of quantum SVM and classical SVM in processing samples from the breast cancer dataset and the wine dataset, Shah et al. found that quantum computation achieves almost as high an accuracy as classical computation while using fewer samples than classical computation, which is sufficient to prove that quantum computation has a strong ability to process complex data [5].

### 1.1.2 Evolutionary and quantum evolutionary algorithms

Evolutionary Algorithms are heuristic optimisation algorithms that simulate natural evolutionary mechanisms. They work by iteratively enhancing solution quality, imitating natural selection and genetic mechanisms. Evolutionary Algorithms are frequently employed to identify the best or nearly best solutions to intricate challenges.

Quantum Evolutionary Algorithms is an optimisation algorithm that combines quantum computing and evolutionary algorithms. It has special quantum mechanisms such as quantum bits, quantum superposition, quantum entanglement and quantum interference, which are unique to quantum computing. The speed and accuracy of these mechanisms in solving certain complex problems can be achieved in a way that cannot be achieved by traditional algorithms.

### 1.2 Aims and Objective

This paper aims to apply the quantum genetic algorithm to solve the traveling salesman problem (TSP). The experiments will assess the potential of this algorithm in tackling NP-hard problems like TSP, comparing its results with those obtained using the traditional genetic algorithm under various conditions. The performance will be evaluated based on criteria such as optimal solution, average optimal solution, optimal number of iterations, average number of iterations for the optimal solution, maximum execution time, and average execution time.

## 2 Theoretical knowledge and related works
### 2.1 Quantum algorithm

Quantum algorithms have in common with classical algorithms in that they both have the basic steps of going through initialisation, obtaining input data, executing and returning results [6]. They differ in that classical computer follow the von Neumann architecture which uses binary operations [7], but quantum computers operate using quantum bits that can perform quantum operations. The development of quantum algorithms has been facilitated by the improvement of quantum computing technology. Wang, J. et al. attempted to simulate quantum Fourier transforms with quantum logic gates and used quantum registers instead of spatial matrices as the storage structure for quantum bits, which helps to increase the speed of execution of quantum algorithms [8]. Shor's algorithm is a classical quantum algorithm that performs the prime factorisation of large integers on a quantum computer. The idea of shor algorithm is to transform factorisation into a periodic problem of finding a certain function, which makes it possible to perform prime factorisation of large integers in polynomial time. Mounica et al. implemented the shor algorithm using 5 quantum bits and successfully factorised the integer 15 [9]. However, the shor algorithm is currently not feasible to break real-world encryption systems such as RSA due to the immaturity of quantum computer technology. Another quantum algorithm is Grover's algorithm. Its main application is searching in unordered databases. For a list containing N elements, Grover's algorithm can find a particular element in about $O(\sqrt{N})$ time complexity, whereas classical algorithms require $O(N)$ time complexity. Khanal et al. have explored Grover's algorithm and amplitude amplification to translate classical logic gates into quantum circuits, leveraging quantum computing properties to potentially reduce the computational power needed to solve certain classical problems [10].

Quantum algorithms also have great potential in the field of machine learning. Gushanskiy et al. described how to use quantum neural networks to solve a number of common image recognition problems, and they contributed to the application of quantum techniques in the framework of image processing [11]. They used Grover's quantum tracking algorithm in retrieving the positions of the image vertices, an algorithm that stores the image in an array of quantum bits that has a faster execution speed than traditional methods. Ablayev et al. demonstrated the use of quantum nearest neighbour algorithms for bifurcation-valued problem classification and compared their execution efficiency with classical algorithms and found that quantum algorithms are approximately a quadratic time faster to execute than traditional algorithms [12]. The Quantum Phase Estimation Algorithm (QPE) is a key component of many more complex quantum algorithms (e.g., Shor's algorithm), which is used to estimate the phase introduced to a quantum state by a quantum gate operation. Ha, J. et al. demonstrated that the One control qubit Quantum Phase Estimation algorithm (OQPE) has better performance and lower error rate than QPE when implemented in a noisy quantum processor [13]. Quantum algorithms in finance can be very useful in helping investors to identify the best portfolios, Upadhyay et al. used variational quantum eigen solver and quantum approximate optimisation algorithm for a company's financial investment for portfolio optimisation, the conclusions given by these two quantum algorithms are highly similar to the models derived fromclassical Markowitz mathematical theory, proving the feasibility of these two algorithms [14].

**2.2 Evolutionary algorithm**
Many real-life complex problems are related to sequential optimisation problems, and evolutionary algorithms are one of the best methods for solving such problems [15]. This is because evolutionary algorithms have a strong ability to balance the quality of the solution as well as the processing time.

The Differential Evolutionary Algorithm, introduced by Storn and Price in 1995, is an evolutionary algorithm designed for addressing global optimization problems. Its optimisation steps include initialisation, mutation, recombination and selection. Its special feature is that it uses the differences of pairs of randomly selected samples in the population to generate offspring. Qu, B. Y. et al. proposed a multi-objective version of the Differential Evolutionary Algorithm to handle multi-objective optimization problems, they also incorporated a technique involving the summation of normalized objective values into the multi-objective Differential Evolutionary Algorithm [16]. This addition significantly enhances the algorithm's efficiency and stability in managing such multi-objective optimization challenges.

Particle swarm optimisation algorithm is a well-known evolutionary algorithm. In this approach, every solution to the problem is treated as a "particle," and each particle is assigned initial velocity and position values. These particles adapt their fitness and velocity in order to search for the global optimal solution based on the best experience of themselves and other particles in the space. Liang, J. J. et al. proposed a multi-objective dynamic multi-swarm particle swarm optimisation algorithm [17]. They applied this algorithm to address a large-scale portfolio optimization problem. This algorithm demonstrated strong performance in terms of both convergence speed and the quality of the optimal solution it produced.

Shi, Y. et al. found that the traditional difference algorithm converges slower than the particle swarm optimisation algorithm, they tried to incorporate the concept of learning from one's own and the group's experience in the particle swarm optimisation algorithm into the difference algorithm, and this improved algorithm combining the advantages of the two algorithms demonstrated better convergence speeds and higher quality solutions [18].

Genetic algorithm is an optimisation search algorithm, which is a classical evolutionary algorithm. It is good at solving optimisation and search problems. Backpack problems can generally be solved by completely traversing the search space using recursive backtracking to find the solution to the problem. However, when the complexity of the problem is too large, the solution space of this method increases exponentially making it impossible to use the recursive backtracking method. Zhao, J. et al. investigated a genetic algorithm based on the implementation of a greedy strategy that can solve the knapsack problem well [19]. In order to reduce the number of iterations, the algorithm does not randomly generate a set of solution sequences when initialising the population as the traditional genetic algorithm do, but instead indicates whether a solution is selected or not in binary form. After comparing the average optimal solution and the average number of iterations of the traditional genetic algorithm and the genetic algorithm based on the greedy strategy respectively, he found that the latter algorithm performed better.

Genetic algorithm can be combined with other optimisation algorithms to improve the performance of the algorithms in solving problems. Liu, B. et al. used the features of ant colony algorithms in parallel processing of information and global search, and the genetic algorithm in enlarging the richness of the solutions to obtain better convergence speed and efficiency in searching for the global optimal solution in dealing with successive optimisation problems, respectively [20]. Genetic algorithm is good at solving some unconstrained optimisation problems, but in real scenarios many problems are constrained. After each iteration, the individuals in the population are updated with new individuals due to crossover and mutation, and some of these new individuals are in the feasible region and some are in the infeasible part. Gao, Y. G. et al. compute the midpoint between individuals in the feasible and infeasible regions, then adjust the parameters used to update the individuals so that the midpoint converges to the feasible individuals and ultimately forces the infeasible individuals to be feasible while keeping the population size unchanged [21]. The new algorithm is tested with a constrained linear programming problem and is shown to improve both the speed of convergence and the accuracy of the solution. Genetic algorithm can be combined with other complex algorithms, and it can play an assisting role in designing Convoltuional Neural Network (CNN) algorithms. Sun, Y. et al. used genetic algorithm to help users with no prior knowledge of the CNN domain to discover and design optimal CNN architectures for dealing with the image classification problem [22]. It was found that this algorithm, which is able to design and tune parameters automatically, showed better performance in terms of both optimal classification accuracy and computational resource usage than traditional algorithms with manually tuned parameters. Genetic algorithm can have different ways of population evolution when dealing with the traveling salesman problem. Wei, G. and Xie, X. first sorted each individual after initialisation according to fitness and then selected the best half of the individuals for crossover and mutation [23]. Finally, the updated individuals and the highly adapted individuals among the old ones were selected and retained to the next generation. This crossover of two well-adapted individuals gives a higher probability of obtaining a better-adapted individual. This evolutionary approach has been tested to find the optimal solution faster.

Genetic programming is an evolutionary algorithm that mimics the evolution of life, and it is very similar to the genetic algorithm for problem solving. However genetic programming does not look for a solution rather it looks for a computer programme. Genetic programming usually stores the results in a tree structure, Suttasupa, Y. et al. compared and studied the encoding methods of genetic programming and they found that Multi-expression Programming has faster convergence than using traditional tree structure encoding [24].

Evolutionary algorithms can also be combined with each other to solve large-scale global optimisation problems. LaTorre, A. and Molina, D. experimented with a multiple offspring sampling framework [25]. This framework combines multiple individual evolutionary algorithms in a particular order, with the initialised population of the latter algorithm being the final population produced by the former. The experimental results showed that ordering multiple evolutionary algorithms in the form of Ascending Participation Order gave the best results.

**2.3 Quantum evolutionary algorithm**
Li, Y. et al. address the multicast routing problem using a quantum-inspired evolutionary algorithm that automatically balances exploration and exploitation operations, compared with the traditional genetic algorithm, its convergence speed and diversity are better with the expansion of network scale [26]. Quantum particle swarm optimization is a quantum evolutionary algorithm that combines quantum theory and optimization algorithm, and its solution space and problem search space are not the same. The probability function of the particle position reflects the state of the particle in the search space, rather than the specific position information of the particle. The process of particles moving to the lowest potential energy point in the field is the process of quantum particle swarm optimization [27].

Hota, A. R. and Pat, A. implemented an adaptive quantum-inspired differential evolution algorithm that adaptively controls mutation and crossover parameters [28]. It shows better convergence results than the traditional difference algorithm when solving 0-1 knapsack problem.

QGA incorporates the concepts of genetic algorithm and quantum computing and is a classical quantum evolutionary algorithm. In traditional genetic algorithm individuals are usually represented as a bit string (0 or 1). However, in QGA, individuals are represented as quantum bits or qubits, and a key property of qubits is that they can exist in a "superposition" state, which means that they can be at 0 and 1 at the same time, a property that allows the QGA to explore multiple solutions in the search space at the same time, which can help to improve the performance of the search.

The QGA distinguishes itself from the traditional genetic algorithm through its approach to updating the population. In the QGA, quantum gates are primarily employed to manipulate quantum superposition states, thereby altering the probability amplitudes associated with the ground state [29].

According to Guo, J. et al., it was argued that the conventional population structure of the QGA establishes a network in which every member is directly connected to every other, resulting in a network with a high clustering coefficient and a short average path length [30]. Although this structure is conducive to information sharing among chromosomes, it may destroy the chromosome diversity and eventually lead the algorithm to fall into a local optimal solution because one chromosome with the highest fitness will be selected as a sample for all chromosomes to evolve each time the population is updated. In response Newman and Watts (1999) improved the population structure of the QGA and introduced the NW network model. The model retains strong connections between randomly selected nodes and adds weak connections between long distance nodes that do not change the clustering coefficients but reduce the average path length in the network. When updating chromosomes, a subpopulation has the probability to select the optimal chromosomes in other subpopulations as its own evolutionary target. The introduction of the NW network model concept allows the QGA to do so with higher execution efficiency and quality of results while maintaining the diversity of the population. Liu, X. et al. proposed three limitations that exist in standard QGA. Firstly, the process of obtaining binary coding by measuring the states of quantum bits on the chromosomes has a great deal of randomness because each quantum chromosome may degrade as the population is optimised. Second, although binary coding is suitable for the knapsack problem and the traveling salesman problem, it is not suitable for problems such as finding the extremes of a function, which require frequent coding and decoding. Third, it is difficult for QGA to accurately determine the magnitude and direction of the angle at which the quantum revolving gate rotates during the process of population evolution [31]. In order to reduce the impact of these problems on the efficiency of the QGA and the accuracy of the results, they use the phase of quantum bits to encode the chromosomes directly and use the gradient information of the fitness function as a variable parameter to calculate the angle of rotation of the revolving gate in the optimisation process.

## 3 Methodology
This part will introduce in detail the methods of implementing QGA in this paper and how to adjust and optimize the operation of QGA according to the traveling salesman problem. In addition, we will introduce the experimental data set and experimental design respectively.

### 3.1 Quantum genetic algorithm
QGA combines the concepts of quantum computing and genetic algorithm, and it enhances the efficiency of problem solving by modelling quantum mechanics. This study implements a simulation of the QGA to operate on a classical computer. Both QGA and traditional genetic algorithm are based on evolutionary ideas with basic operations. However, the existence of quantum properties makes the encoding method of QGA, the parallelism of the search space of the solution, the updating method of the solution, the determinism of the solution and the search strategy different from the traditional genetic algorithm. In addition, since the QGA in this paper does not run on a real quantum computer but is simulated on a classical computer, its performance may be worse than that of the traditional genetic algorithm in solving some real problems.

QGA is different from the basic operation of traditional genetic algorithm because of its quantum mechanism. The main steps of QGA include quantum population initialisation, quantum measurement, fitness assessment, selection, quantum crossover, quantum mutation, quantum chromosome update, and iteration.

### 3.1.1 Quantum population initialization

Quantum population initialisation is the beginning of a QGA. A quantum population is randomly generated, and each quantum chromosome constituting the population consists of a certain number of quantum bits. A quantum bit differs from a classical bit in that it can be in a superposition of two quantum states at the same time. The $\alpha|0\rangle$ in the equation $|\psi\rangle = \alpha|0\rangle + \beta|1\rangle$ denotes the spin-down state and $\beta|1\rangle$ denotes the spin-up state. A single quantum bit can contain both $|0\rangle$ and $|1\rangle$ information [32]. Similar to traditional genetic algorithm QGA also requires setting the population size and chromosome length. Where the population size is adjusted appropriately with the size of the problem. Chromosome length is equal to the number of cities to be visited.

Unlike the classical genetic algorithm, which employs qubit coding, the QGA uses a completely different coding technique. For the traveling salesman problem, since each gene may have multiple states, it is necessary to adopt multi-qubit coding [33]. Each gene on a chromosome is made up of a number of quantum bits, in this paper the relationship between the number of bits n of a quantum bit and the number N of visiting cities is $n=\log_2 N$ (n rounded down). After quantum measurements these quantum bits collapse to the classical state, at which point the corresponding city sequence number can be obtained by calculating the decimal value of each gene representation. There are i strings of quantum bits on the genes of a quantum chromosome, it can represent $2^i$ simultaneous states.

$$|\psi\rangle \leftarrow \begin{pmatrix} \alpha_1 & \alpha_2 & \alpha_3 & \cdots\cdots & \alpha_i \\ \beta_1 & \beta_2 & \beta_3 & \cdots\cdots & \beta_i \end{pmatrix}. \tag{1}$$

Fig. 1 illustrates the representation of quantum superposition states.

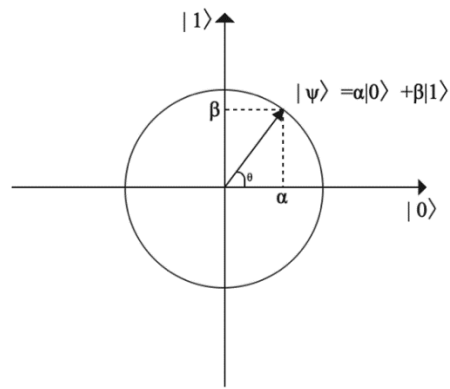

**Fig. 1** Representation of quantum superposition states

This paper initialises the quantum population by setting the values of all quantum bit amplitudes such that all quantum superposition states have the same probability of being expressed. A quantum bit can be put into a superposition state by applying the Hadamard matrix. A quantum superposition state is achieved by multiplying the Hadamard matrix with the $|0\rangle$ vector.

$$H * |0\rangle = \begin{pmatrix} \frac{1}{\sqrt{2}} & \frac{1}{\sqrt{2}} \\ \frac{1}{\sqrt{2}} & \frac{-1}{\sqrt{2}} \end{pmatrix} * \begin{pmatrix} 1 \\ 0 \end{pmatrix}. \tag{2}$$

The phase angle $\theta$ is a random value between $0 \sim \frac{\pi}{2}$, and take $\theta$ as a parameter of the rotation matrix U.

$$U = \begin{pmatrix} \cos\theta & -\sin\theta \\ \sin\theta & \cos\theta \end{pmatrix}. \tag{3}$$

Finally multiplying the vector of quantum superposition states with the quantum rotation matrix gives the quantum bit state represented by the amplitude $(\alpha_i, \beta_i)$.

$$\begin{pmatrix} \alpha_i \\ \beta_i \end{pmatrix} = U*(H*|0\rangle) = \begin{pmatrix} \cos\theta & -\sin\theta \\ \sin\theta & \cos\theta \end{pmatrix} * \begin{pmatrix} \frac{1}{\sqrt{2}} & \frac{1}{\sqrt{2}} \\ \frac{1}{\sqrt{2}} & \frac{-1}{\sqrt{2}} \end{pmatrix}. \tag{4}$$

In a quantum population, chromosomes are made up of genes, genes are made up of quantum bits, and the state of a quantum bit is determined by the amplitude $(\alpha_i, \beta_i)$.

**3.1.2 Quantum measurement**

The purpose of quantum measurement is to collapse the quantum bits in the superposition state into specific values so that the fitness value of each chromosome can be calculated. In this paper, we determine the state of a quantum bit by comparing the values of $\alpha^2$ and $\beta^2$ (the probability that the quantum state $|\psi\rangle$ collapses to $|0\rangle$ and $|1\rangle$ ).

**3.1.3 Fitness evaluation**

The QGA uses quantum bit coding, so it is necessary to first convert the quantum bit values obtained from the quantum measurements, which are expressed in binary, to be expressed in decimal, thus obtaining the city serial number. However, this method is difficult to avoid that there will be duplicates of randomly generated city serial numbers in a chromosome, which is not in accordance with the rules of TSP. Therefore, in this paper, we add a step before calculating the fitness value to find out the duplicated and missing cities in all chromosomes and replace the duplicated city serial numbers with the missing ones. Although this method will lose some efficiency of the algorithm execution, but it can ensure that the solutions obtained subsequently are legitimate.

The QGA expresses the fitness of the quantum chromosome in terms of the sum of the distances of the paths visiting all the cities. The individual with the smallest fitness value in each generation is defined as the best individual.

**3.1.4 Selection**

In this paper, we try to use elite selection as a method of selection operation for QGA. Firstly, the whole population is sorted according to the fitness, and then n individuals with better fitness are used to directly replace the individuals with poorer fitness. This selection method is conducive to retaining the optimal solution in the population and improving the convergence speed.

**3.1.5 Crossover**

When using a QGA for this problem, the direct use of standard quantum crossover operations may lead to illegal solutions, such as the presence of duplicate cities in a chromosome. This is due to the fact that CNOT gate operations may result in certain genes (cities) appearing multiple times in a sub-chromosome. To address this issue, the quantum crossover operation of the QGA used in this paper is similar to the crossover operation of the traditional genetic algorithm, but it requires that the quantum chromosome is first converted into a decimal representation. This has the advantage that we can operate on them more directly and ensure that the result is still a legitimate TSP path. The crossover operations are performed on populations that are better adapted overall after elite selection. These individuals with excellent fitness have a greater likelihood of crossing each other to produce better adapted individuals. Fig. 2 illustrates the quantum chromosome crossover process.

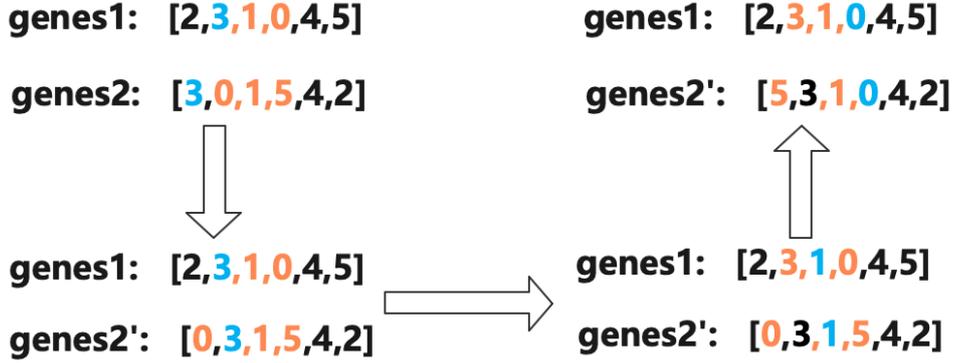

**Fig. 2** Chromosome crossover process

**3.1.6 Quantum mutation**

Quantum mutation is similar to the mutation operation of the traditional genetic algorithm in that both are designed to increase population diversity and jump out of local optima. The difference is the object of operation. QGA operates on quantum bits. In this paper, the conventional quantum mutation operation is used to change the state of quantum bits. Changing the state of quantum bits from $|0\rangle$ to $|1\rangle$ or from $|1\rangle$ to $|0\rangle$ is achieved by exchanging the values of quantum bit amplitudes $\alpha$ and $\beta$. In this paper, probability values are set for chromosomes and quantum bits respectively to decide whether mutation is performed or not. The efficiency of the algorithm can be optimised by adjusting the mutation probability according to the characteristics of the specific problem.

**3.1.7 Quantum chromosome update**

Updating quantum chromosomes is a crucial step in QGA and approaching the states of all quantum chromosomes in the population towards the optimal solution generated in each generation can better approximate the global optimal solution. The quantum revolving gate is an actuator to achieve quantum chromosome updating [33].

In this paper, we use quantum revolving gate to implement updates to chromosomes. The quantum revolving gate is defined as:

$$U(t) = \begin{pmatrix} \cos(\theta\prime) & -\sin(\theta\prime) \\ \sin(\theta\prime) & \cos(\theta\prime) \end{pmatrix}. \tag{5}$$

The quantum revolving gate updates the state of each pair of quantum bit amplitudes in the population towards the optimal individual in a process:

$$\begin{pmatrix} \alpha_i^{t+1} \\ \beta_i^{t+1} \end{pmatrix} = \begin{pmatrix} \cos\theta\prime & -\sin\theta\prime \\ \sin\theta\prime & \cos\theta\prime \end{pmatrix} \begin{pmatrix} \alpha_i^t \\ \beta_i^t \end{pmatrix}. \tag{6}$$

$\theta'$=s($\alpha_i$,$\beta_i$) * $\Delta\theta_i$, where s($\alpha_i$,$\beta_i$) is the direction of rotation and $\Delta\theta_i$ is the rotation value. The execution schematic of the revolving gate is shown in Fig. 3 and Fig. 4.

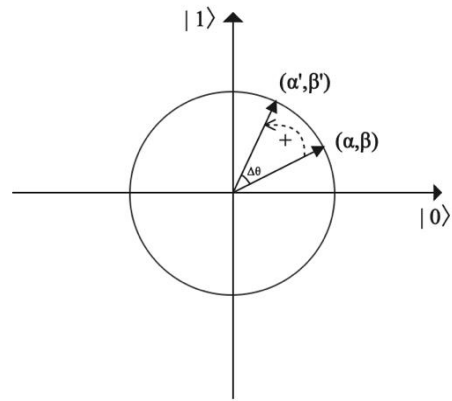

**Fig. 3** Rotation of quantum bits from state 0 to state 1

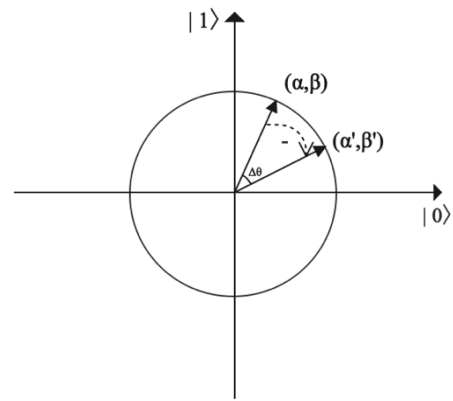

**Fig. 4** Rotation of quantum bits from state 1 to state 0

If the quantum bit of the chromosome that serves as the evolutionary sample is 1 and the quantum bit of the chromosome currently to be updated is 0, it is rotated in the positive direction. If the quantum bit of the chromosome that serves as the evolutionary sample is 0 and the quantum bit of the chromosome currently to be updated is 1, it is rotated in a negative direction.

The value of $\Delta\theta_i$ is the angular step size of the rotation, which has a significant effect on the efficiency of the algorithm. A short angular step is good for accurate search but will reduce the search speed, and a too long angular step may lead to excessive dispersion of the results or converge to the local optimal solution quickly [32]. In this paper, we adopt a dynamic adjustment of the angular step size strategy, which gradually reduces the rotation angle as the number of iterations increases. The purpose is to explore the solution space more at the beginning of the algorithm, while the solution optimisation can be carried out more finely at the later stage of the algorithm.

**3.1.8 Iteration**
After selection, crossover, quantum mutation and quantum chromosome updating the better adapted individuals of the quantum population will enter the next generation and the best individuals will be recorded as the current optimal solution and will be the model for the evolution of other individuals in the next generation.

**3.2 Experiment**

**3.2.1 Optimisation issues and TSP**

An optimisation problem is a problem of finding the best solution under some constraints. These problems generally consist of a fitness function, decision variables, and constraints. TSP is a classical combinatorial optimisation problem, which can be seen as an extension of the Hamilton circle problem. Given a set of city coordinates, a traveller needs to find the shortest path that allows him to start from a city, visit each city exactly once and return to the starting city.

We selected the TSP as a benchmark to evaluate the performance of both the traditional genetic algorithm and the QGA for the following reasons: firstly, TSP has the property of finding an optimal solution under specific constraints, which occurs in many real-world applications (e.g., logistics, urban public transport path planning problems, etc.), and so the study of TSP can easily be generalised to other domains. Second, TSP is an NP-hard problem, which means that the time required for computation may grow exponentially as the number of cities increases. Due to this computational complexity, it is important to find efficient algorithmic solutions for TSP.

**3.2.2 Dataset**

The TSP dataset selected for this paper is derived from the data in the international standard dataset TSPLIB. This dataset contains city location coordinates and corresponding optimal solutions for symmetric TSP instances. The library contains datasets of real-world geographic locations, which provides researchers with the opportunity to test their algorithms in a variety of real-world scenarios. In addition, test cases ranging from few to many and from simple to complex exist in TSPLIB, which can widely satisfy the need to test algorithms with different performances.

Four datasets of different sizes were chosen for the experiment, including Burma 14, Ulysses 16, Bayg 29 and Att 48. Burma 14 and Ulysses 16 are small datasets. The advantage of choosing small datasets is their shorter data processing and execution time, which can quickly help us test whether the algorithms can be executed during the implementation process and make it easier to identify potential problems during debugging. Bayg 29 and Att 48 have a larger and more complex number of use cases, which means that they have better representativeness and diversity, and they are closer to the scale of real-world problems, which makes them more valuable for research.

**3.2.3 Experimental design**

In this paper, genetic algorithm and QGA are used to solve different data sets. The tests were all run on the same classical computer. For the accuracy and credibility of the experiment, we solved each data set 10 times with each algorithm. During the experiment, in order to find the parameters that can make both algorithms play a better performance at the same time, we constantly adjust the parameters of population size, crossover probability and mutation probability. During the adjustment, it is found that with the increase of the number of cities, appropriately increasing the population size and crossover probability can increase the diversity of the population and make the algorithm have a larger solution space to explore, thus improving the possibility of finding the optimal solution. However, too large population and too high crossover probability will greatly increase the execution time of the algorithm. The setting of mutation probability should be very careful. If the value is too small, it will be difficult for the algorithm to skip the local optimal solution; if the value is too large, it will destroy some good solutions in the current population, and even make the algorithm become the same as the random search algorithm, resulting in the performance of the algorithm.

To accommodate different dataset sizes, we made adjustments to the algorithm parameters. Specifically, for the Burma 14 and Ulysses 16 datasets, we set the population size for both the traditional genetic algorithm and the QGA to 80, with a crossover probability of 0.7 and a mutation probability of 0.3. When dealing with larger datasets like Bayg 29 and Att 48, we increased the population size to 120 and set the crossover probability to 0.9, while retaining the mutation probability at 0.3. Additionally, we fine-tuned the quantum revolving gate's update angle, which is calculated as delta_theta = 0.01 * (generation_max / (generation + 1)). Here, 'generation_max' represents the maximum number of iterations, and 'generation' denotes the current number of iterations. These parameters were thoroughly tested and chosen through repeated experiments. It was ensured that setting these parameters led to improved experimental outcomes for both the traditional genetic algorithm and the QGA.

Fig. 5 displays the flowchart of the traditional genetic algorithm.

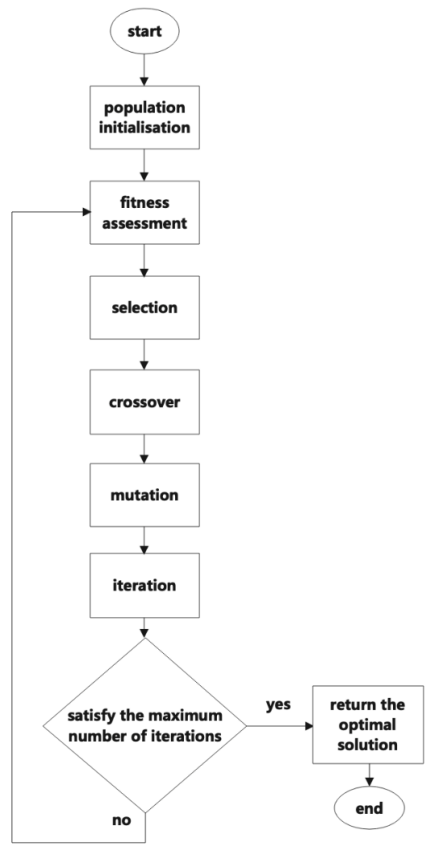

**Fig. 5** Flowchart of traditional genetic algorithm

Fig. 6 displays the flowchart of the QGA.

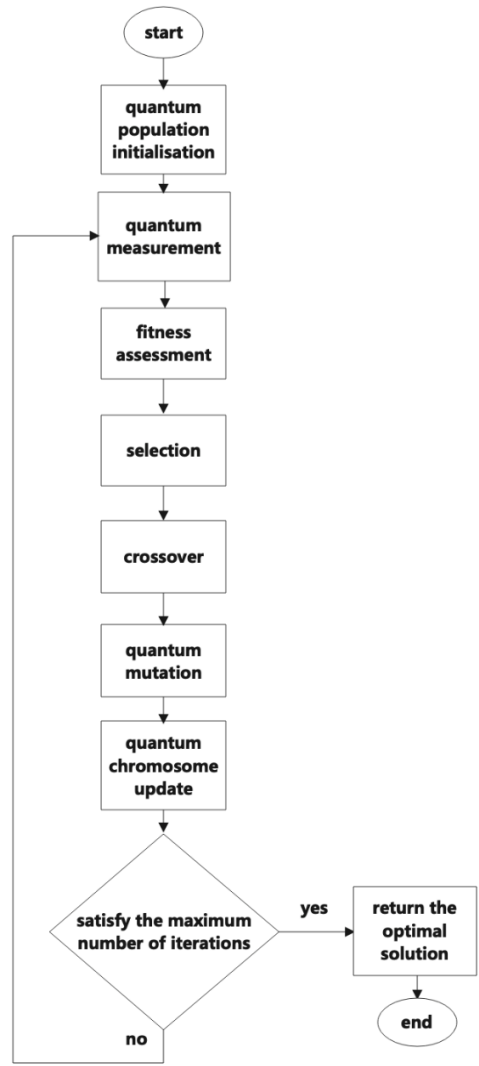

**Fig. 6** Flowchart of QGA

## 4 Results and discussions

The performance evaluation of both the traditional genetic algorithm and the QGA is based on several key metrics, including the optimal solution, average optimal solution, optimal number of iterations, average number of iterations for the optimal solution, maximum execution time, and average execution time. The results are reported for both the best-case scenario and the average case. The average case values are computed after running each dataset ten times.

In this section, the experimental outcomes, fitness curves, and generated paths for each dataset are presented, analyzed, and discussed. The fitness value of each chromosome is determined based on the length of the path it represents. As the algorithm progresses, the fitness value decreases, reflecting the optimization process.

### 4.1 Burma 14 dataset

**Table 1** Results when processing the Burma 14 dataset

|     | Optimal solution | Average optimal solution | Optimal number of iterations | Average number of iterations of the optimal solution | Maximum execution time(s) | Average execution time(s) |
| --- | --- | --- | --- | --- | --- | --- |
| GA  | 30.879 | 31.236 | 45 | 48.20 | 0.310 | 0.319 |
| QGA | 30.879 | 31.607 | 68 | 107.7 | 39.88 | 41.815 |

The fitness curves for GA and QGA when processing the Burma 14 dataset are shown in Fig. 7 and Fig. 8, respectively.

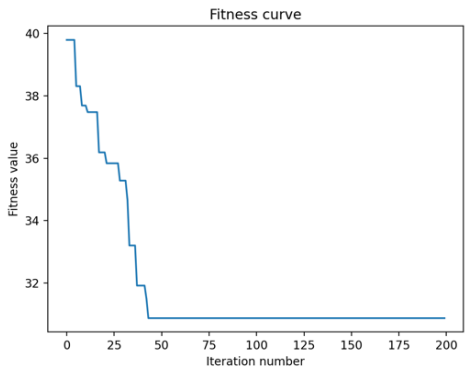
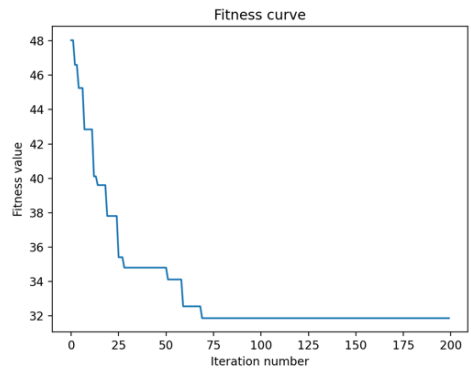

Fig. 7 Fitness curve of GA             Fig. 8 Fitness curve of QGA

The paths generated by GA and QGA when processing the Burma 14 dataset are shown in Fig. 9 and Fig. 10, respectively.

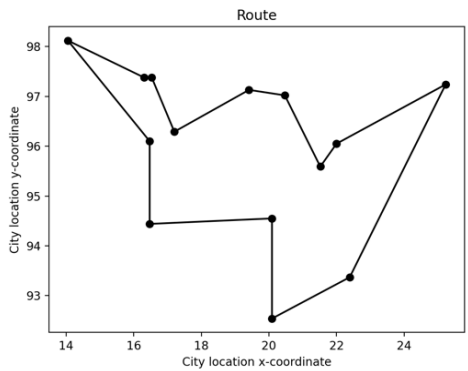
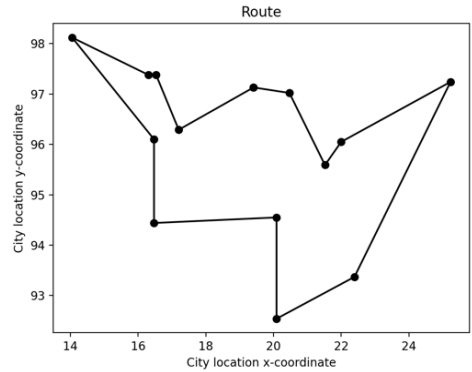

Fig. 9 The path generated by GA             Fig. 10 The path generated by QGA

Based on the data results and the generated path diagrams, it's evident that both the genetic algorithm and the QGA successfully discover the global optimal solution and yield the same path when dealing with the Burma 14 dataset. This shows that both algorithms satisfy the basic requirements for solving the TSP. Furthermore, this paper provides a comparative analysis of the performance of the genetic algorithm and the QGA. The average optimal solutions of these two algorithms have a very small difference and both are very close to the optimal solution, which indicates that both algorithms are very capable and stable for searching the optimal solution for the Burma 14 dataset. However, it's worth noting that in comparison to the conventional genetic algorithm, the QGA needs more execution time and iterations to discover the best answer. There appears to be algorithm stability because the number of optimal iterations and the average number of iterations for the conventional genetic algorithm are similar. In contrast the QGA has a much higher average number of iterations than the optimal number of iterations, which indicates that its search process is not stable. This may be due to the fact that the superposition state of quantum bits allows the search space to be explored more extensively thus causing the algorithm to require more iterations to converge. Analysis of fitness curves during the execution of both algorithms reveals that the traditional genetic algorithm exhibits a more consistent trend in updating the optimal solution. It consistently produces superior individuals roughly every five generations. The QGA, on the other hand, will appear for 25 consecutive generations without updating the optimal solution, which is easy to fall into the local optimal solution. This reflects that the traditional genetic algorithm has a stronger ability to get out of the local optimal solution. The traditional genetic algorithm and QGA show a clear difference in execution time, and the long execution time of the QGA may be due to the fact that the simulation of quantum phenomena on classical computers (quantum gate operations, superposition state processing, encoding and decoding) needs to be more complex and time-consuming than the conventional operations in the traditional genetic algorithm.

## 4.2 Ulysses 16 dataset

**Table 2** Results when processing the Ulysses 16 dataset

|  | Optimal solution | Average optimal solution | Optimal number of iterations | Average number of iterations of the optimal solution | Maximum execution time(s) | Average execution time(s) |
| --- | --- | --- | --- | --- | --- | --- |
| GA | 73.987 | 74.520 | 45 | 83.1 | 0.35 | 0.358 |
| QGA | 73.999 | 75.139 | 40 | 78.7 | 58.64 | 63.104 |

The fitness curves for GA and QGA when processing the Ulysses 16 dataset are shown in Fig. 11 and Fig. 12, respectively.

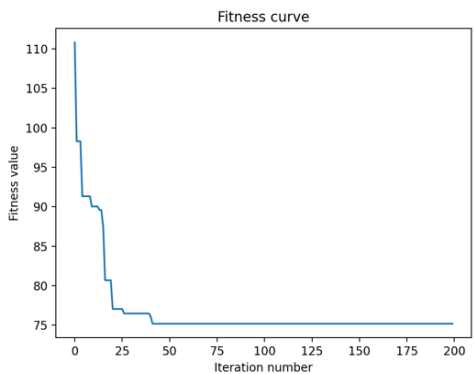 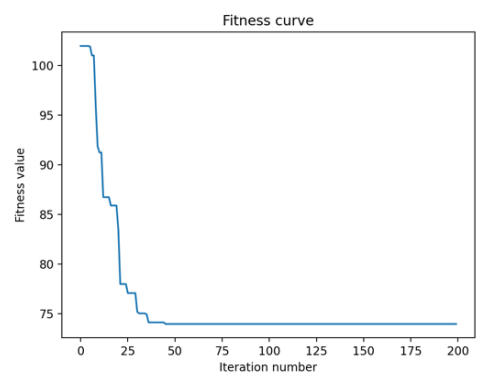

Fig. 11 Fitness curve of GA            Fig. 12 Fitness curve of QGA

The paths generated by GA and QGA when processing the Ulysses 16 dataset are shown in Fig. 13 and Fig. 14, respectively.

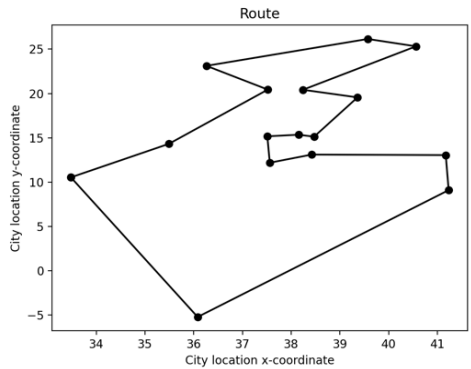 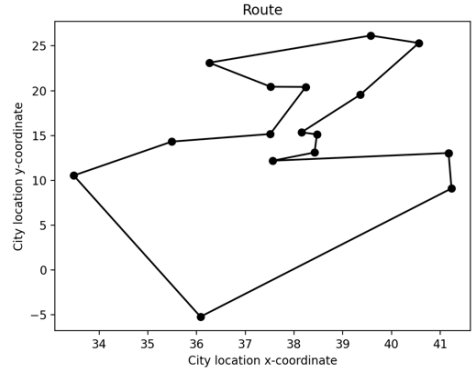

**Fig. 13** The path generated by GA      **Fig. 14** The path generated by QGA

For the Ulysses 16 dataset, it's observed that the traditional genetic algorithm produces a slightly better optimal solution compared to the QGA. Moreover, the average optimal solution from the traditional genetic algorithm is very close to the optimal solution, indicating the high stability of the search results delivered by the traditional genetic algorithm. Compared with the traditional genetic algorithm, the gap between the average optimal solution and the optimal solution of the QGA is a little larger, showing that the stability of the algorithm decreases slightly as the problem size increases. It's worth noting that, although the disparity between the optimal solutions of the traditional genetic algorithm and the QGA is not substantial, there is a significant difference in the paths they plan. This may be due to the fact that there are a large number of possible paths in the TSP, which means that there may be multiple paths with costs close to the optimal solution, but their structures are very different. In addition, the fact that the traditional genetic algorithm and the QGA rely on different strategies for searching for paths (e.g., selection, crossover, and mutation are used in the traditional genetic algorithm, while the QGA introduce quantum manipulation and superposition of states) also leads to large differences in the structure of paths found by the two algorithms. This reflects the fact that diversity of solutions is equally important, sometimes the final solutions differ greatly in their presentation, but they may both be well adapted.

Surprisingly, the QGA requires fewer minimum iterations and fewer average iterations than the traditional genetic algorithm to find the optimal solution on this dataset. It demonstrated better convergence performance with the same population size, crossover probability and mutation probability than when dealing with the Burma 14 dataset. When analysed in conjunction with the data results for the average optimal solution, this could be the result of the QGA falling earlier into some local optimal solution that is close to the global optimal solution. This may also be due to the specific use case of the Ulysses 16 dataset, which makes it easy for the QGA to fall into a certain local optimal solution. The difference between the average number of iterations and the minimum number of iterations of the traditional genetic algorithm when dealing with the Ulysses 16 dataset is larger than when dealing with the Burma 14 dataset, reflecting a decrease in the stability of the number of iterations of the traditional genetic algorithm as the problem size increases.

The fitness curves of the two algorithms show a very similar trend of rapid convergence at the beginning of the algorithm, being able to jump out of the local optimum after it has been encountered, and finally finding the optimal solution. Where the QGA encounters the local optimal solution earlier than the traditional genetic algorithm.

The execution time of the QGA remains longer than the traditional genetic algorithm. Combining the execution times of the two algorithms for the Burma 14 dataset shows that there is no significant increase in the execution time of the traditional genetic algorithm. In contrast the QGA has a more significant increase in execution time, indicating that its computational speed decreases faster as the problem size increases.

**4.3 Bayg 29 dataset**

**Table 3** Results when processing the Bayg 29 dataset

|  | Optimal solution | Average optimal solution | Optimal number of iterations | Average number of iterations of the optimal solution | Maximum execution time(s) | Average execution time(s) |
| --- | --- | --- | --- | --- | --- | --- |
| GA | 9104.36 | 9276.437 | 136 | 150.9 | 2.55 | 2.803 |
| QGA | 9866.51 | 11076.557 | 453 | 508.21 | 410 | 486.89 |

The fitness curves for GA and QGA when processing the Bayg 29 dataset are shown in Fig. 15 and Fig. 16, respectively.

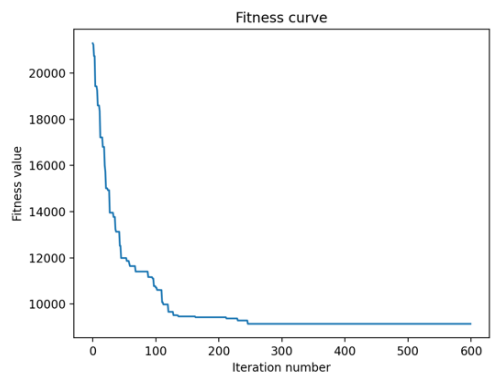 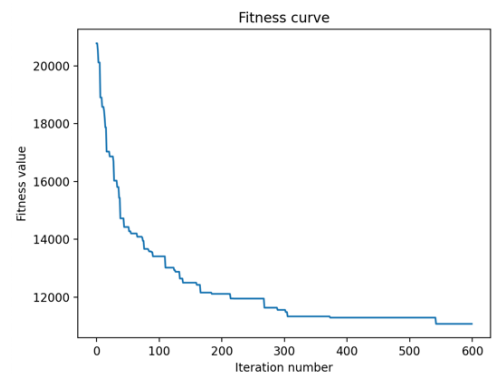

**Fig. 15** Fitness curve of GA         **Fig. 16** Fitness curve of QGA

The paths generated by GA and QGA when processing the Bayg 29 dataset are shown in Fig. 17 and Fig. 18, respectively.

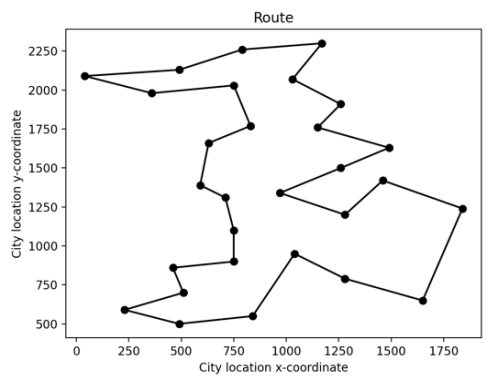 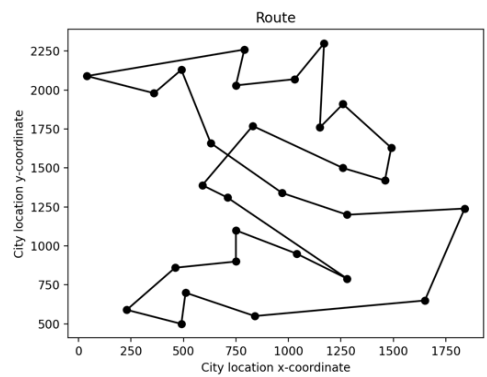

**Fig. 17** The path generated by GA         **Fig. 18** The path generated by QGA

The known optimal solution for the Bayg 29 dataset is 9074.15, in this experiment the optimal solution obtained by the traditional genetic algorithm is much closer to this value and the error between the average optimal solution and the optimal solution is only less than 2%, which is almost the same as the error that occurs in the two relatively small datasets of Burma 14 and Ulysses 16. This represents the fact that the expansion of the dataset does not have a significant effect on the stability of the optimal solutions computed by the traditional genetic algorithm. The optimal solution and the path generated by the QGA can be analysed to show that its results for the Bayg 29 dataset are not very satisfactory. Its average error between the optimal solution and the optimal solution reaches 11%, which is a significant increase compared to the previous two datasets of 6% and 2%, which indicates that the quality of the solutions computed by the QGA becomes more unstable with the expansion of the problem size.

The fitness curves of these two algorithms when dealing with the Bayg 29 dataset are significantly different compared to the fitness curves when dealing with the Burma 14 dataset as well as the Ulysses 16 dataset, and we can see that there is a significant slowdown in the rate of convergence, which decreases more rapidly in the case of the QGA.

As the TSP dataset and population size increased further, the QGA did not continue to show a trend of fewer iterations when dealing with the Ulysses 16 dataset than when dealing with the Burma 14 dataset, but rather increased substantially. The increase in population size and the increase in crossover probability allow the QGA to enhance the space in which it explores solutions, thereby increasing the diversity and computational complexity of the solutions. In addition, QGA may experience longer plateau periods during optimisation, during which the quality of the solutions does not improve significantly, because the algorithm may need to spend more time exploring new and evolutionary potential solution regions. These factors lead to slower convergence of the algorithm and require longer execution time.

**4.4 Att 48 dataset**

Table 4 Results when processing the Att 48 dataset

|  | Optimal solution | Average optimal solution | Optimal number of iterations | Average number of iterations of the optimal solution | Maximum execution time(s) | Average execution time(s) |
|---|---|---|---|---|---|---|
| GA | 33915.24 | 35094.923 | 286 | 376.4 | 11.72 | 12.396 |
| QGA | 42159.381 | 46621.861 | 510 | 536 | 916.25 | 974.33 |

The fitness curves for GA and QGA when processing the Att 48 dataset are shown in Fig. 19 and Fig. 20, respectively.

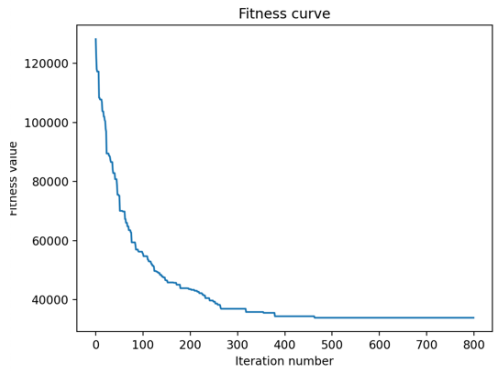 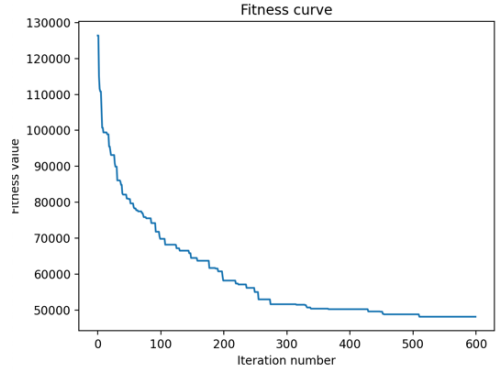

Fig. 19 Fitness curve of GA    Fig. 20 Fitness curve of QGA

The paths generated by GA and QGA when processing the Att 48 dataset are shown in Fig. 21 and Fig. 22, respectively.

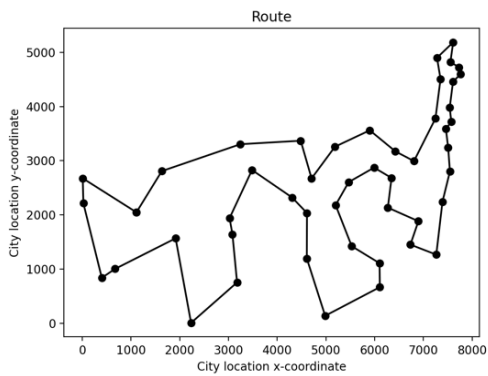 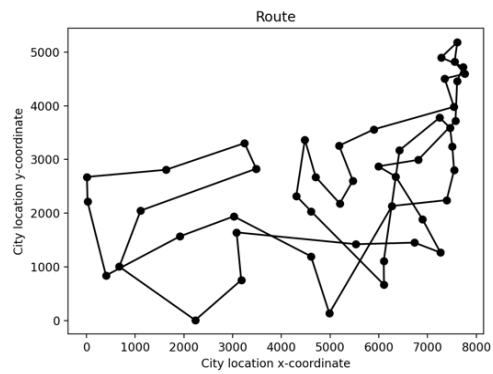

Fig. 21 The path generated by GA    Fig. 22 The path generated by QGA

Att 48 is the dataset with the most and most complex city coordinates in this experiment. The inclusion of this dataset serves the purpose of assessing the performance of both the traditional genetic algorithm and the QGA in solving complex problems, pushing the limits of these algorithms. The known optimal solution for the Att 48 dataset is 33522. As per the experimental results, it is evident that the traditional genetic algorithm comes closer to the optimal solution. On the other hand, the optimal solution achieved by the QGA indicates that it is significantly trapped in a local optimum. The substantial gap between the average optimal solution and the known optimal solution of the Att 48 dataset clearly demonstrates that the current experimental parameter settings and algorithm performance of the QGA are insufficient to handle this specific dataset.

Comparing the path diagrams generated by the traditional genetic algorithm and the QGA, we can find that there are a lot of paths that cross each other in the path results generated by the QGA, which is obviously not in line with the optimal solution. However, the traditional genetic algorithm does not find the optimal solution, but its generated paths do not cross each other, which seems to be reasonable and valuable.

The maximum number of iterations was set to 600 when dealing with the Att 48 dataset due to the long execution time required by the QGA and the slow rate of convergence. The fitness curve shows that the

QGA converges more slowly and is prone to long plateau periods as the algorithm runs making it difficult to find solutions with evolutionary potential.

The gap between the average number of iterations and the minimum number of iterations of the QGA is smaller than that of the traditional genetic algorithm, which can be analysed in combination with the fitness curve to conclude that the limit of its algorithmic performance when dealing with the Att 48 dataset is roughly stable at around 500 generations. Compared with the traditional genetic algorithm, the comprehensive performance of the QGA decreases more as the problem size increases.

**5 Conclusion**

The aim of this paper is to implement the basic functionality of the QGA and then use it to solve the classical travelling salesman problem and compare its performance difference with that of the traditional genetic algorithm. For QGA we implemented quantum population initialisation, quantum measurement, fitness assessment, selection, crossover, quantum mutation, quantum chromosome update and iteration. In our experiments we chose the classical city coordinates dataset from the international standard dataset TSPLIB as the experimental use case and experimented with the traditional genetic algorithm and QGA sequentially using datasets ranging from simple to complex. We collected the optimal solutions of the two algorithms, the number of iterations and the execution time as the reference indexes for evaluating the performance of the algorithms. In addition, we plot the fitness curves and the final paths generated by both algorithms when dealing with different datasets. By analysing the results of these experimental data, we draw conclusions. Both the traditional genetic algorithm and the QGA achieve more satisfactory results for these datasets, especially when dealing with smaller sized datasets. Overall, the traditional genetic algorithm performed better on most of the test cases. It showed better optimal solutions, number of iterations, execution time, and convergence speed. Multiple experiments on the same dataset also verified that the traditional genetic algorithm has better stability than the QGA. As the number of TSP cities increases, we observe that the performance of both the traditional genetic algorithm and the QGA decreases, there is a fall into local optimal solutions, and there is an increase in the number of iterations and the execution time of the algorithm. Among them, the performance of the QGA decreases more significantly, especially when dealing with larger scale datasets there is a long plateau period, resulting in poor convergence efficiency of the algorithm. Although the optimal performance and stability of the QGA is weaker than that of the traditional genetic algorithm, in the process of adjusting the experimental parameters, we found that it can find an optimal solution similar to the traditional genetic algorithm in a smaller population size, and this potential feature can give the QGA an advantage when dealing with large datasets.

According to our experimental results, both traditional genetic algorithm and QGA can achieve good results when dealing with datasets with smaller data sizes, but traditional genetic algorithm seems to be a better choice when dealing with larger datasets. In our future research we will further optimise and adjust the parameters of both the traditional genetic algorithm and the QGA to improve their performance on larger scale TSP datasets. Since the population initialisation method and quantum bit encoding approach for the QGA in this paper may cause duplicate cities in the chromosomes, the paths need to be checked and repaired during decoding. This means that the quantum population cannot be updated perfectly at each iteration. This is much less efficient than the traditional genetic algorithm to update the solution, so in future research we will design a more suitable method to initialise the quantum population. In this paper the QGA is simulated and run on a classical computer, the simulation of quantum operations requires a large number of computational resources, which increases the execution time. With the advancement of quantum computing technology, the performance of quantum algorithms may be greatly improved if they can be run on a real quantum computer. In addition to the traveling salesman problem, we will apply these two algorithms to other optimisation problems in the future to assess their generality and execution efficiency, and we will try to combine the strategies used by traditional genetic algorithm and QGA to obtain better solutions and performance.